\documentclass[prx,twocolumn,amsmath,amssymb,notitlepage]{revtex4-1}
\usepackage{tikz}
\usepackage{circuitikz}
\usetikzlibrary{arrows,shapes.gates.logic.US,shapes.gates.logic.IEC,calc}
\usepackage{graphicx}
\usepackage{amssymb}
\usepackage{amsfonts}
\usepackage{amsmath}
\usepackage{amsbsy}
\usepackage{natbib}
\usepackage{comment}
\usepackage{amssymb}
\usepackage{color}
\usepackage{float}
\DeclareMathAlphabet\mathbfcal{OMS}{cmsy}{b}{n}
\DeclareMathAlphabet{\mathpzc}{OT1}{pzc}{m}{it}
\usepackage[colorlinks=true, urlcolor=blue, citecolor=blue, linkcolor=blue]{hyperref}
\usepackage{bm}
\usepackage{bm}
\usepackage{bbold}

\renewcommand{\vec}[1]{\mbox{\boldmath$\mathrm{#1}$}}
\let\sb=_ \catcode`\_=\active \def_#1{\ensuremath \sb{\rm#1}}

\renewcommand{\vec}[1]{\mbox{\boldmath$\mathrm{#1}$}}
\newcommand{\be}{\begin{equation}}
\newcommand{\ee}{\end{equation}}
\newcommand{\ben}{\begin{eqnarray}}
\newcommand{\een}{\end{eqnarray}}

\begin{document}


\title{Skyrmion lattice hosted in synthetic antiferromagnets and helix modes}

\author{X.-G. Wang$^{1}$, L. Chotorlishvili$^{2,3}$, G. Tatara$^{4,5}$, A. Dyrda\l$^{6}$, Guang-hua Guo$^{1}$,  V. K. Dugaev$^2$, J. Barna\'s$^{6,7}$, S.\,S.\,P. Parkin$^{8}$, and A. Ernst$^{8,9}$}
\address{$^1$ School of Physics and Electronics, Central South University, Changsha 410083, China \\
    $^2$ Department of Physics and Medical Engineering, Rzesz\'ow University of Technology, 35-959 Rzesz\'ow, Poland\\
    $^3$ Faculty of Mathematics and Natural Sciences, Tbilisi State University, Chavchavadze av.3, 0128 Tbilisi\\
	$^{4}$ RIKEN Center for Emergent Matter Science, 2-1 Hirosawa, Wako, Saitama 351-0198, Japan\\
    $^{5}$ RIKEN Cluster for Pioneering Research, 2-1 Hirosawa, Wako, Saitama, 351-0198 Japan\\
	$^6$ Faculty of Physics, Adam Mickiewicz University, ul. Uniwersytetu Pozna\'nskiego 2, 61-614 Pozna\'n, Poland\\
    $^7$ Institute of Molecular Physics, Polish Academy of Sciences, ul. M. Smoluchowskiego 17, 60-179 Pozna\'{n}, Poland\\
    $^8$ Max Planck Institute of Microstructure Physics, Weinberg 2, D-06120 Halle, Germany\\
	$^{9}$ Institute for Theoretical Physics, Johannes Kepler University, Altenberger Stra\ss e 69, 4040 Linz, Austria}

\date{\today}
\begin{abstract}
  Thin ferromagnetic films can possess unconventional magnetic properties, opening a new road for using them in spintronic technologies. In the present work exploiting three different methods, we comprehensively analyze phason excitations of a skyrmion lattice in synthetic antiferromagnets. To analyze phason excitations of the skyrmion lattice, we have constructed an analytical model based on three coupled helices and found a linear gapless mode.  Micromagnetic simulations also support this result. Moreover, a similar result has been
achieved within the rigid skyrmion lattice model based on the coupled Thiele's equations, when the coupling between skyrmions in different layers of the synthetic antiferromagnetic is comparable to or larger than the intralayer coupling.  In addition, we also consider the orbital angular momentum and spin pumping current associated with phason excitations. Due to the gapless excitations in the case of skyrmion lattice, the pumping current is nonzero for the arbitrary frequency of pumping microwaves. In the case of individual skyrmions, no current is pumped when microwave frequency is inside the gap of the spectrum of individual skyrmions.
\end{abstract}

\maketitle

\section{Introduction}
There is currently a great interest in two-dimensional topological
solitons (skyrmions) and in ordered skyrmion lattices, known also as
skyrmion crystals (SkX)
\cite{barton2020magnetic,schroers1995bogomol,seki2012observation,wilson2014chiral,white2014electric,derras2018quantum,
  haldar2018first,leonov2015multiply,
  psaroudaki2017quantum,van2013magnetic,rohart2016path,samoilenka2017gauged,battye2013isospinning,
  jennings2014broken,tsesses2018optical}.  It is well established that
the dominant interaction leading to skyrmion formation is the
Dzyaloshinskii–Moriya (DM) coupling that occurs in magnets with no
spacial inversion symmetry. This coupling lowers the ground state
energy of the system and thus stabilizes the skyrmion magnetic
textures. Formation of SkXs in thin films is energetically more
favourable than formation of individual skyrmions.  A key problem is a
search for materials hosting SkXs. In what follows, we will explore
the formation of SkXs and also their dynamical properties in a synthetic
antiferromagnet (SAF), i.e. in a system consisting of two
ferromagnetic layers coupled antiferromagnetically. Individual
skyrmions in such materials were investigated very recently
\cite{legrand2020room}.

Before proceeding to the main objectives of this paper, we briefly
recall the key features of the magnonic spectrum of ferromagnets with
individual skyrmions and with SkXs. The dynamical properties of
individual skyrmions are studied in
Refs. \cite{PhysRevB.90.094423,PhysRevB.97.064403,PhysRevB.89.024415}.
It was shown that the spectrum of low-energy excitations in a
ferromagnetic layer hosting a single static skyrmion includes a magnon
mode with an energy gap \cite{PhysRevB.90.094423}. The dispersion of
this mode is $\omega(p)=\omega_0(a/R)^2+\omega_0 a^2 p^2$, where
$\omega_0$ is the stiffness frequency related to the exchange
interaction, $R$ is the skyrmion radius, $a$ the lattice parameter, and $p$ is the radial
momentum. Inside the frequency gap, $\omega_0(a/R)^2$, there appear
two localized states, related to the bound skyrmion-magnon breathing
and quadrupole modes \cite{PhysRevB.90.094423}. Since the energy of
a system with a single skyrmion does not depend on the position of the
skyrmion, there
is also a zero energy mode associated with the skyrmion drift: a skyrmion can
move as a massless particle in a gauge field.

In the case of an SkX, the continuous symmetry of the system is
broken. Nevertheless, in-plane translations of the SkX lattice as a
whole do not change the system's energy, which leads to a gapless magnon mode
corresponding to the deformation waves in the SkX lattice. Naturally,
these magnon modes can be associated with the gapless Nambu-Goldstone
excitations \cite{PhysRevD.90.025010}, which appear at the phase
transition breaking the initial symmetry of Lagrangian. This problem
has been discussed in a number of publications
\cite{muhlbauer2009skyrmion,tatara2014phasons,PhysRevLett.120.067201,PhysRevB.82.134427,PhysRevB.84.214433,
  PhysRevB.87.100402,PhysRevB.92.140415,
  PhysRevB.93.174429,PhysRevB.100.100408,PhysRevLett.108.017601,
  PhysRevLett.120.077202,rozsa2020spin,seki2020propagation,li2020bimeron}.

 A central problem in the magnonic spintronics is the rectification and
control of the magnonic spin current \cite{wang2018electric,
  PhysRevB.92.174411, PhysRevX.6.031012}.
Direction of this current can be switched by an
external magnetic field \cite{PhysRevX.6.031012}. However, the
magnetic field increases the gap in magnon spectrum and thus reduces
the number of magnons contributing to the magnonic spin
current. In the present work, we show that the gapless spectrum of SkX
allows switching of the spin current without reducing its
magnitude. Apart from this, due to swirling of the magnetization texture in the SkX,
the net spin current in both layers of a SAF is nonzero, while
it vanishes in the SAF without SkX. Thus,  SkX in a SAF may serve as
a unique platform for manipulating spin currents in  spintronic devices.

Generally, the free energy of a ferromagnetic system, as a function of
the unit vector $\bf m$ pointing along the magnetization, can be
written in the form \cite{muhlbauer2009skyrmion}:
\begin{eqnarray}\label{2Free energy}
F_{SkX}\left[\vec{m}(\textbf{r})\right]=\int\left[A_{\rm{ex}}\left(\vec{\nabla}\vec{m}(\textbf{r})\right)^{2}-\mu_0 M_s m_z H_z\right.\nonumber\\
\left. + a_m\vec{m}^2(\textbf{r}) + b_m\vec{m}^4(\textbf{r}) + E_{DM}\right]d^2\vec{r},
\end{eqnarray}
where the first and second terms correspond to the exchange and Zeeman
energy, respectively, where $M_s$ is  the saturation magnetization
and $A_{\rm{ex}}$ is the exchange stiffness parameter. The
	last term,
	$E_{DM}=D [(m_z \frac{d m_x}{dx} - m_x \frac{dm_z}{dx}) + (m_z \frac{d m_y}{dy} - m_y \frac{dm_z}{dy})]$, stands for the interfacial DM
	energy and breaks symmetry in the \textbf{z} direction. The free energy of an SkX also includes the Ginzburg–Landau
terms, $a_m\vec{m}^2(\textbf{r})$ and $b_m\vec{m}^4(\textbf{r})$, that generally
are essential for stabilization of the magnetization. The  Ginzburg–Landau  energy, Eq.(\ref{2Free energy}),
 is valid  close to the Curie temperature $T_c$, and was used to argue stabilization of the skyrmion lattice structure by a quartic term $\sim m^4$~\cite{muhlbauer2009skyrmion}. Thus, this approach accounts for the emergence of SkX  near Tc. Until recently, SkX has been shown to appear in various temperature regimes due to different stabilization mechanisms. In this paper, however, we will not deal with  the stabilization mechanisms of the SKX, so we assume that the three-helix state (see Eq. (3) in the following section)  is a good approximation.

It has been shown that a SkX can be considered as a superposition of
three coupled helices~\cite{PhysRevB.105.024422,PhysRevB.103.104408, 
PhysRevB.103.094402}. 
At temperatures below a critical temperature of transition to the trivial magnetic phase, and for intermediate magnetic fields, the description based on three magnetic helices is well-justified~\cite{PhysRevB.103.094402}.
A ferromagnetic layer with a single helix and
with coupled helices was studied in~\cite{PhysRevB.84.214433}.  It was
shown there that the quadratic part of the free energy for a single
helix can be diagonalized exactly, and below the critical value of
$a_m$, $a_m<D^2/(4A_{ex})$, the ground state is a single helix with the
energy $\epsilon (k)=a_m+A_{ex}k^2-Dk$ which is minimized for
$k=D/2A_{ex}\equiv Q$.  Small excitations from the ground state have
been considered 
in terms of the Euler-Lagrange equations for the Lagrangian function
$L=L_B-U$, where $L_B$ stands for the relevant Berry phase term,
which plays the role of kinetic energy, while $U$ takes into account
energy due to magnetization deviations from the ground state.
A simple analytical formula was found for spin waves propagation along
the helix, with the corresponding dispersion relation
$\omega(p)=(2A_{ex}\gamma M_s)\sqrt{Q^2p^2+p^4}$,
where $\gamma$ is the gyromagnetic ratio. The spectrum is gapless and
linear in the low energy limit, i.e. for $p\ll Q$.
To study coupled helices, an external magnetic field was assumed, that
induces a finite uniform magnetization. Owing to this, the quartic
term in the free energy could be rewritten as an effective cubic term,
which couples the three helices~\cite{PhysRevB.84.214433}.  Two types
of modes were discovered: (i) longitudinal waves,
$\omega_{l}=A_{ex}\gamma M_s\sqrt{3Q^2p_{in}^2+2p^4_{out}}$,
associated with the displacement of the SkX parallel to the in-plane
component of the wave vector ${\bf p}_{in}=(p_x,p_y,0)$, and (ii)
transverse waves,
$\omega_{t}=A_{ex}\gamma M_s\sqrt{Q^2p_{in}^2+2p^4_{out}}$, with the
displacement along
$\hat{\textbf{z}}\times\textbf{p}_{in}=(-p_y,p_x,0)$, where
$\hat{\textbf{z}}$ is a unit vector along the axis $z$, and $p_{out}$
is the out-of-plane ($z$) component of the wave vector. For details
see Ref.~\onlinecite{PhysRevB.84.214433}.

\begin{figure}
  \includegraphics[width=\columnwidth]{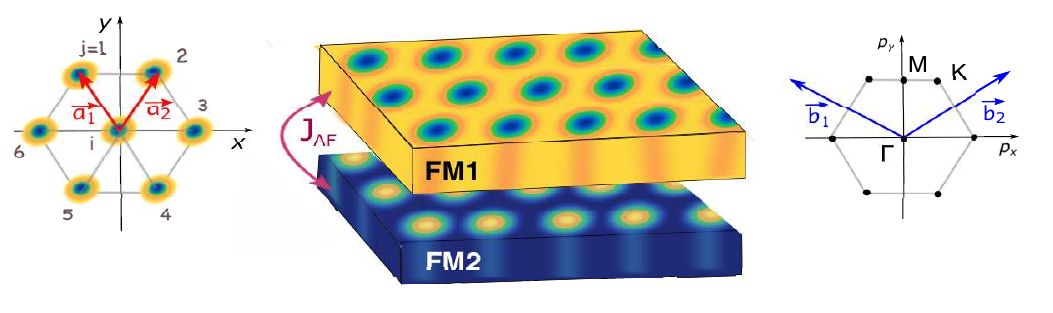}
  \caption{\label{model} (a) Schematic of antiferromagnetic
    skyrmions in a SAF, where two ferromagnetic layers, FM1 and
    FM2, are coupled {\it via} an antiferromagnetic interlayer
    exchange coupling $J_{AF}$. The DM interaction stabilizes
    the Skyrmion structure. Each layer exerts a bias field on
    the neighboring layer. Skyrmions form a regular triangular
    lattice (b) with the SkX lattice vectors ${\bf a}_1$ and
    ${\bf a}_2$. The SkX lattice parameter is
    $a_{SkX} = |{\bf a}_1| = |{\bf a}_2|\equiv r$, where $r$ is
    the inter-skyrmion distance. (c) Corresponding
    Brillouin zone.}
\end{figure}

In the present work, we explore the phason excitation spectrum in the
SAF (see Fig. \ref{model}).
Phasons are excitations corresponding to a phase degree of freedom of the collective structures, proposed originally in \cite{PhysRevB.18.6245}.
In-plane translations of the SkX lattice conserve the system's energy, while continuous symmetry within the SkX is broken.
The gapless sliding phason modes in the SkX lattice are equivalent to the gapless Nambu-Goldstone excitations mentioned above. The idea is general for any structure with
periodicity.
In fact, for a periodic structure  $\sim\sin(kx)$ along the $x$ direction and with a wave length $k$, shifting the  coordinate center by a time dependent position $X(t)$ leads to a dynamic phase $\varphi(t)\equiv kX(t)$, as $\sin(k(x-X(t))=\sin(kx -\varphi(t))$.
The phason excitations can also be described by a standard perturbative expansion of fluctuations \cite{PhysRevB.84.214433}. However, the approach used in \cite{tatara2014phasons} for description of the phason modes is useful for  physical interpretation of the structure dynamics, and also for comparison of analytical results and experimental
observations. We note that the nontrivial magnetic
texture of antiferromagnetic skyrmions promotes a non-vanishing
topological spin Hall effect \cite{PhysRevLett.121.097204}. The SAF is
composed of nanometer-thick ferromagnetic layers, which are coupled
antiferromagnetically through a non-magnetic spacer layer.  The
coupling mechanism arises from the Ruderman-Kittel-Kasuya-Yoshida
(RKKY) interaction. The heavy-metal-ferromagnet interfaces lead to the
DM interaction in the SAF.  Skyrmions in SAFs are stabilized by the DM
interaction and bias magnetic
fields from other neighbouring layers. We analyze the magnetic
dynamics using different approaches.

The work is organized as follows: In Sec.~\textbf{II}, we generalize
the method based on coupled helices to study phason excitation in the
SAF case. In Sec.~\textbf{III}, we present results of micromagnetic simulations. The obtained results are
consistent with those based on the coupled helices model. In turn, in
Sec.~\textbf{IV}, we present a Thiele's equation approach for a single
ferromagnetic layer and for two layers coupled antiferromagnetically
(SAF system). We also discuss the differences between the methods.
The Thiele's equations treat skyrmions as rigid objects, neglecting
thereby magnetization dynamics inside the skyrmion magnetic texture.  The
SkX in a SAF is modeled in this approach as a lattice of interacting
individual skyrmions. In comparison, the coupled helices model is
relevant for the strongly interacting and correlated phase, where the
concept of individual skyrmions is irrelevant. Despite this, we
show that the approach based on Thiele's equation leads to results
which are consistent with those obtained by other techniques. A summary
and final conclusions are presented in Sec.~\textbf{V}.

\section{Spectrum of phason excitations in a SAF}

\newcommand{\ellv}{\bm{\ell}}
\newcommand{\mv}{\bm{m}}
\newcommand{\nv}{\bm{n}}
\newcommand{\kv}{\bm{k}}
\newcommand{\zv}{\bm{z}}
\newcommand{\rv}{\bm{r}}
\newcommand{\lt}{\left}
\newcommand{\rt}{\right}
\newcommand{\nnr}{\nonumber\\}

It is known that ferromagnetic or antiferromagnetic ordering is not possible in one or two-dimensional Heisen
berg systems with finite-range interaction \cite{PhysRevLett.17.1133} (see also generalization of this statement to some long-range interaction
models \cite{PhysRevLett.87.137203}). The system considered in this paper consists of two ferromagnetic layers of a finite (though small) thickness, that are coupled antiferromagnetically across a thin nonmagnetic spacing layer. Accordingly, the Mermin-Wagner theorem is not applicable to the system under consideration, and we assume that magnetic order is not destroyed by thermal fluctuations. One should also note that even in pure 2D case fluctuations of the order parameter are growing
logarithmically at large distances, which makes it possible to neglect this effect in real finite samples.

Based on the free energy given by Eq.~(\ref{2Free energy}), one can
describe the skyrmion lattice in a ferromagnetic layer as a
superposition of three helices \cite{muhlbauer2009skyrmion}.  The
corresponding low energy excitations were studied in
Refs. \cite{PhysRevB.84.214433,tatara2014phasons} and a gapless mode
of phason excitations with the dispersion quadratic in the wave vector
was identified in the absence of pinning.  In this representation, the
magnetization vector can be parameterized as
${\bf m} = m_z\hat{\bf{z}}+ \sum_{\mu =a,b,c}\bf{m}_{\mu }$, where
$m_z$ is a uniform component induced by an external magnetic field,
while $\bf{m}_{\mu }$ is the spatial profile of the three helices,
\begin{align}
{\bf m}_{\mu }
  &= m_h \lt( \beta_\mu  \hat{\bf{k}}_\mu  +\sqrt{1-\beta_\mu ^2}\, \bf{n}_{\mu }\rt),
  \label{Mvi}
\end{align}
for $\mu =a,b,c$.  Here $m_h$ is the helix amplitude and $\nv_{\mu }$
represent the helices with wave vectors $\kv_\mu $,
\begin{align}
  {\bf n}_{\mu } &= \hat{\bf z} \cos({\bf k}_\mu \cdot {\bf r}+\varphi_\mu ) +(\hat{{\bf k}}_\mu \times \hat{\bf z}) \sin({\bf k}_\mu \cdot {\bf r}+\varphi_\mu ), \label{nvidef}
\end{align}
and $\hat{{\bf k}}_\mu$ is a unit vector along ${\bf k}_\mu$.  One
choice of the vectors ${\bf k}_\mu $ is: ${\bf k}_a=k\, (1,0,0)$,
${\bf k}_b=k\, (-\frac{1}{2},\frac{\sqrt{3}}{2},0)$ and
${\bf k}_c=k\, (-\frac{1}{2},-\frac{\sqrt{3}}{2},0)$.  In turn, the
three variables $\varphi_\mu $ describe phases of the helices, while
$\beta_\mu $ represent massive excitations.  The low energy
excitations are described by the phason variables defined as
$\varphi_+ \equiv \frac{1}{2}(\varphi_a+\varphi_b)-\varphi_c$ and
$\varphi_- \equiv \frac{1}{2\sqrt{3}}(\varphi_a-\varphi_b)$, and the
corresponding low-energy Lagrangian was shown to have the form
\cite{tatara2014phasons}
\begin{align}
L &=  \int d^2{\bf r} \lt\{ g\, ( \varphi_+\dot{\varphi}_- - \varphi_- \dot{\varphi}_+) +\frac{m_\varphi}{2}(\dot{\varphi}_+^2 + \dot{\varphi}_-^2)
\rt. \nnr & \lt.
 -\tilde{A}_{ex} \lt[ (\nabla\varphi_+)^2+(\nabla\varphi_-)^2 \rt]
   \rt\} , \label{Lphi2}
\end{align}
where $g$ is a constant proportional to the topological charge of the
skyrmion, $\tilde{A}_{ex}$ is a constant proportional to the exchange
parameter $A_{ex}$, and $m_\varphi \propto 1/(\tilde{A}_{ex}D)$ is a
mass term arising from the $\beta_\mu$ modes.  Equation~(\ref{Lphi2})
leads to the excitation mode which is quadratic in the wave vector
${\bf p}$, $\omega (p)\propto \tilde{A}_{ex}p^2$. We note that the description based on Eq.(2) is the simplest approximation, validity of which, however,  is  confirmed numerically. More general and accurate descriptions are based on elliptical (deformed) spirals and also include higher order harmonics. In this paper, however, we limit the description to helices described by Eq. (2).

The excitation mode becomes significantly changed in the case of two
ferromagnetic layers coupled antiferromagnetically.  This change
appears due to the dynamics of the dominant antiferromagnetic
component of the two-layers magnetization.  This component experiences
fluctuations of the ferromagnetic component of the two layers.  We
note that this effect is well-known for antiferromagnets in general.
Let us consider the antiferromagnetic coupling between the two
ferromagnetic layers, labeled with the index $i =1$ (FM1 layer) and
$i =2$ (FM2 layer), see Fig.~\ref{model},
\begin{align}
H_{AF}=J_{AF} \int d^2{\bf r}\, ({\bf m}_{1}\cdot {\bf m}_{2}),
\end{align}
where $ J_{AF} $ is the inter-layer coupling constant.
Defining the antiferromagnetic moment ${\bf n}$  and the ferromagnetic moment $\ellv $  as
\begin{align}
 {\bf m}_{1}/M_s&= \nv+\ellv , & {\bf m}_{2}/M_s= -{\bf n}+\ellv ,
\end{align}
the antiferromagnetic coupling can be rewritten as,
$H_{AF}=J_{AF} \int d^2{\bf r}\, (-{\bf n}^2+\ellv ^2 )$.  The spin
dynamics of the system is described by the spin Berry's phase term in
the Lagrangian,
$L_{\rm B}=M_s\int d^2{\bf r} \sum_{i =1,2}\cos\theta_i\,
\dot{\phi}_i$,
written in terms of polar coordinates.  Defining
$\tilde{\bf m}_2\equiv-{\bf m}_2$ (with polar coordinates
$\tilde{\theta}_2,\tilde{\phi}_2$) this term may be written as
$L_{\rm B}=M_s\int d^2{\bf r}\, (\cos\theta_1\, \dot{\phi}_1
-\cos\tilde{\theta}_2\, \dot{\tilde{\phi}}_2 ) $,
which reduces to
$L_{\rm B}=M_s\int d^2{\bf r}\; \delta {\bf m} \cdot ({\bf m}_1 \times
\dot{\bf m}_1) $
in the lowest order in
$\delta {\bf m}\equiv {\bf m}_1-\tilde{\bf m}_2$.  The spin Berry's
phase term, expressed by ${\bf n}$ and $\ellv $ (assuming small
$\ell$, i.e., large $J_{AF}$), then reads:
\begin{align}
L_{\rm B}&= -2M_s\int d^2{\bf r}\; \ellv \cdot ({\bf n}\times\dot{\bf n}) ,
\end{align}
instead of the topological term for the case of single-layer (Eq. (22)
of Ref. \cite{tatara2014phasons}).  Integrating out the $\ellv$
variable and neglecting the spatial derivatives of $\ellv $, one
obtains the kinetic term for $\nv$
\begin{align}
L_{\rm B}&=
 \frac{M_s^2}{J_{AF}}
\int d^2{\bf r}\; \dot{\bf n}^2. \label{nellLag}
\end{align}
Using
$\int d^2{\bf r}\; \dot{\bf n}^2=\int d^2{\bf r}\sum_\mu
\left(\dot{\varphi}_\mu^2+\dot{\beta}_\mu^2\right)$,
the phason kinetic part of the Lagrangian is
\begin{align}
L_{\rm B}&=  \frac{M_s^2}{3J_{AF}}\int d^2{\bf r}\,
( \dot{\varphi}_+^2+ \dot{\varphi}_-^2 ), \label{neelLag}
\end{align}
and the total phason Lagrangian for the SAF, without the topological
term of a single layer (Eq. (\ref{Lphi2})), can be written as
\begin{align}
L &=  \int d^2{\bf r} \lt\{ \frac{\tilde{m}_\varphi}{2}(\dot{\varphi}_+^2 + \dot{\varphi}_-^2)
 -\tilde{A}_{ex} [(\nabla\varphi_+)^2+(\nabla\varphi_-)^2]\rt\}, \label{LphiSAF}
\end{align}
where $\tilde{m}_\varphi\equiv {m}_\varphi+\frac{2M_s^2}{3J_{AF}}$.
The phason dispersion derived from this Lagrangian is massless and
linear, $\omega (p)\propto \tilde{A}_{ex} p$, in the absence of
pinning.  Micromagnetic simulations (Section III) support the result
of the gapless linear mode.

Applying an {\it ac} magnetic field along the in-plane direction, one
can pump spin current into the substrate layer.  The pumped current is
proportional to ${\bf m}\times \dot{\bf m}$ in the phason picture.  As
discussed in Ref.~\cite{tatara2014phasons} (Eq.~(65)), the in-plane
field couples to the massive excitation modes, $\beta_\mu$, and not
directly to the phason variables. Thus, one can write
$\dot{\bf m}_\mu =m_{\rm h}\dot{\beta}_\mu \hat{\bf k}_\mu $. After
averaging over space, the oscillating components vanish and one
obtains
${\bf m}\times \dot{\bf m}=\sum_{\mu \mu^\prime } {\bf m}_\mu \times
\dot{\bf m}_{\mu^\prime }=m_{\rm h}^2 \sum_{\mu \mu^\prime } (\hat{\bf
  k}_\mu\times\hat{\bf k}_{\mu^\prime })
\beta_\mu\dot{\beta}_{\mu^\prime }$.
As
$(\hat{\bf k}_\mu\times\hat{\bf k}_{\mu^\prime }) \parallel \hat{\bf
  z}$
for $\mu\neq {\mu^\prime }$, the pumped spin current is polarized
along the $z$-direction.  The spatial correlation of the $\beta$-modes
is determined by both gapless and gapful modes, $\omega^\pm ( p)$,
with the energies $\omega^-(p) \propto p^2$ and
$\omega^+(p) \simeq A_{ex}Q^2 +O(p^2)$ (Eq. (66) of
Ref. \cite{tatara2014phasons}). The correlation length of the pumping
is determined by the Gilbert damping constant.

\section{Micromagnetic simulations}

The skyrmion generation and its collective dynamics in a SAF is
governed by the Landau-Lifshitz-Gilbert (LLG) equation,
\begin{equation}
\displaystyle \frac{\partial \vec{M}_i}{\partial t} = - \gamma \vec{M}_i \times \vec{H}_{\mathrm{\rm eff},i} + \frac{\alpha}{M_{s,i}} \vec{M}_i \times \frac{\partial \vec{M}_i}{\partial t}
\label{LLG}
\end{equation}
for the top ($ i = 1$) and bottom ($i= 2 $) ferromagnetic layers in
the SAF. Here, $ \vec{M}_i = M_s \vec{m}_i $ ($ M_s$ denotes the
saturation magnetization), and $ \alpha $ is the phenomenological
Gilbert damping constant. The total effective field
$ \vec{H}_{eff,i} $ exerted on the $ i$-th layer reads:
$\vec{H}_{eff,i}=-\frac{\delta F_{SkX}}{\delta
  \textbf{m}_i}-\vec{H}_{couple,i}$,
where $F_{SkX}$ is given by Eq.~(1),
$ \vec{H}_{couple,i} = \frac{J_{AF}}{\mu_0 M_{s,i} t_i} \vec{m}_j $ is
the bias field exerted by the second layer, $ t_i $ is the $ i $-th
layer thickness, and $ j \ne i $.  The influence of the out-of-plane
magnetic anisotropy and dipole-dipole interaction
is not taken into account in the present description.

In numerical calculations we assume the following parameters:
$ A_{ex} = 10 \;\rm{pJ/m}$, $ D_m = 0.2\; \rm{mJ/m^2}$,
$M_s = 1.2 \;\rm{A/m} $, $ J_{AF} = 0.23\; \rm{mJ/m}^2 $, and
ferromagnetic layer thickness $ t_p = 3\; \rm{nm} $. The bias magnetic
field $ H_z = 100\, $mT is used for stabilization of the skyrmion
structure. The size of the ferromagnetic layers is
$ 6000 \times 120 \times 3\; \rm{nm}^3 $, which is discretized by the
cell size $ 3 \times 3 \times 3\; \rm{nm}^3 $.

\subsection{Single ferromagnetic layer}
\begin{figure}
  \includegraphics[width=\columnwidth]{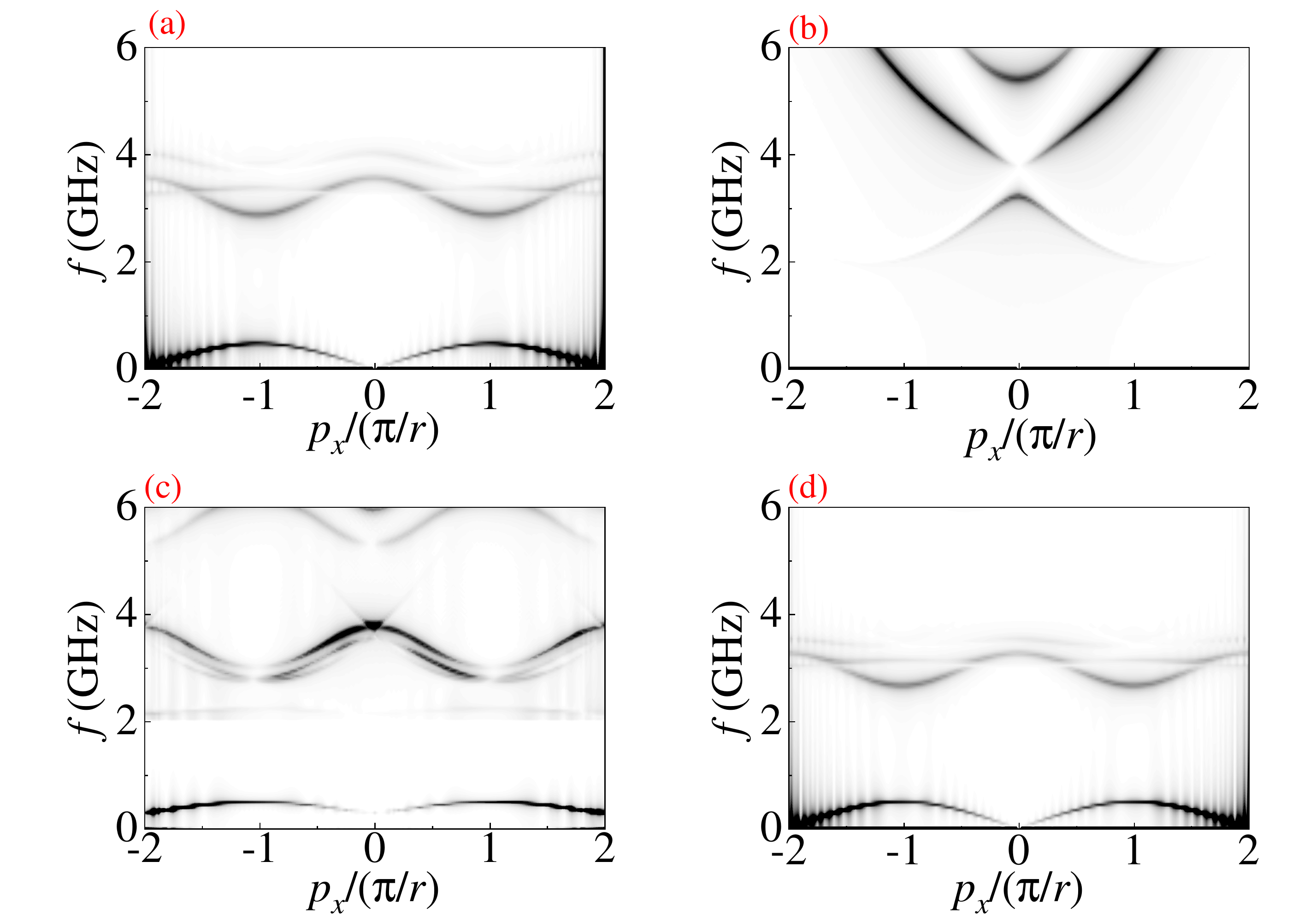}
  \caption{\label{dispersion}  (a) Low-temperature spectrum of magnetic
    oscillations in the single ferromagnetic layer with
    one-dimensional skyrmion lattice. (b)  The corresponding spectrum in the absence of skyrmions.
    (c) The same as in (a)   for  the external magnetic field
    field $ -20\, \rm{mT} $  applied near the skyrmion center
    (radius of this range is 15 nm).
     (d) The same as in (a) but for nonzero Ginzburg-Landau parameters, $ a_m = 1 \times 10^5 $ J/m$ ^3 $ and
    $ b_m = 5 \times 10^4 $ J/m$ ^3 $. These parameters describe spin wave spectrum in the vicinity of Curie temperature. General features of the spectrum are similar to those in (a) except the frequencies are in general  slightly smaller.
    The frequency $f =\omega /(2\pi )$ is shown as a function
    of $p_x$, and the period of 1D SkX is $ r = 61.8 $ nm.
    The Brillouin zone
    boundaries in (a,c,d) are given by $ p_x = \pm n \pi/r $.  }
\end{figure}

For clarity reasons, we analyze first the skyrmion dynamics in a
single magnetic layer, where the theory based on the model of coupled
helices predicts gapless excitations. For simplicity we focus here on
the excitations in a one-dimensional SkX.
The corresponding low-temperature spectrum of magnetization dynamics is presented in
Fig.~\ref{dispersion}(a), where the frequency $\omega /(2\pi )$ is
shown as a function of $p_x$.  Periodicity of the SkX, with the period $r$
($r$ is the distance between skyrmions), is clearly visible in the
corresponding band structure, see Fig.~\ref{dispersion}(a).
%
The gapless excitations near $ p_x=0 $ correspond to the collective
SkX mode, already discussed above.  To emphasize the gapless character
of the collective phason excitations, we also calculated the
excitation spectrum in the magnetic layer without the SkX, where the
excitations are gaped, see Fig.~\ref{dispersion}(b).
Spectrum of magnetic excitations for temperatures close to the Curie-temperature is described by specific  nonzero values of the Ginzburg-Landau parameters, see Fig.~\ref{dispersion}(d). These  parameters are related to the system's  temperature and  Curie temperature, for details see Ref.~\onlinecite{muhlbauer2009skyrmion} and the corresponding supplementary material. The corresponding spectrum is qualitatively similar to that in Fig.~\ref{dispersion}(a), except the corresponding frequencies are slightly smaller, compare Figs.~\ref{dispersion}(a,d).
However, when confining the skyrmions through a pinning potential, this mode
shifts upward, see Fig.~\ref{dispersion}(c).  All these results show,
that the micromagnetic calculations lead to the results, which are
qualitatively consistent with those obtained in the model based on
three coupled helices.
\begin{figure}
  \includegraphics[width=\columnwidth]{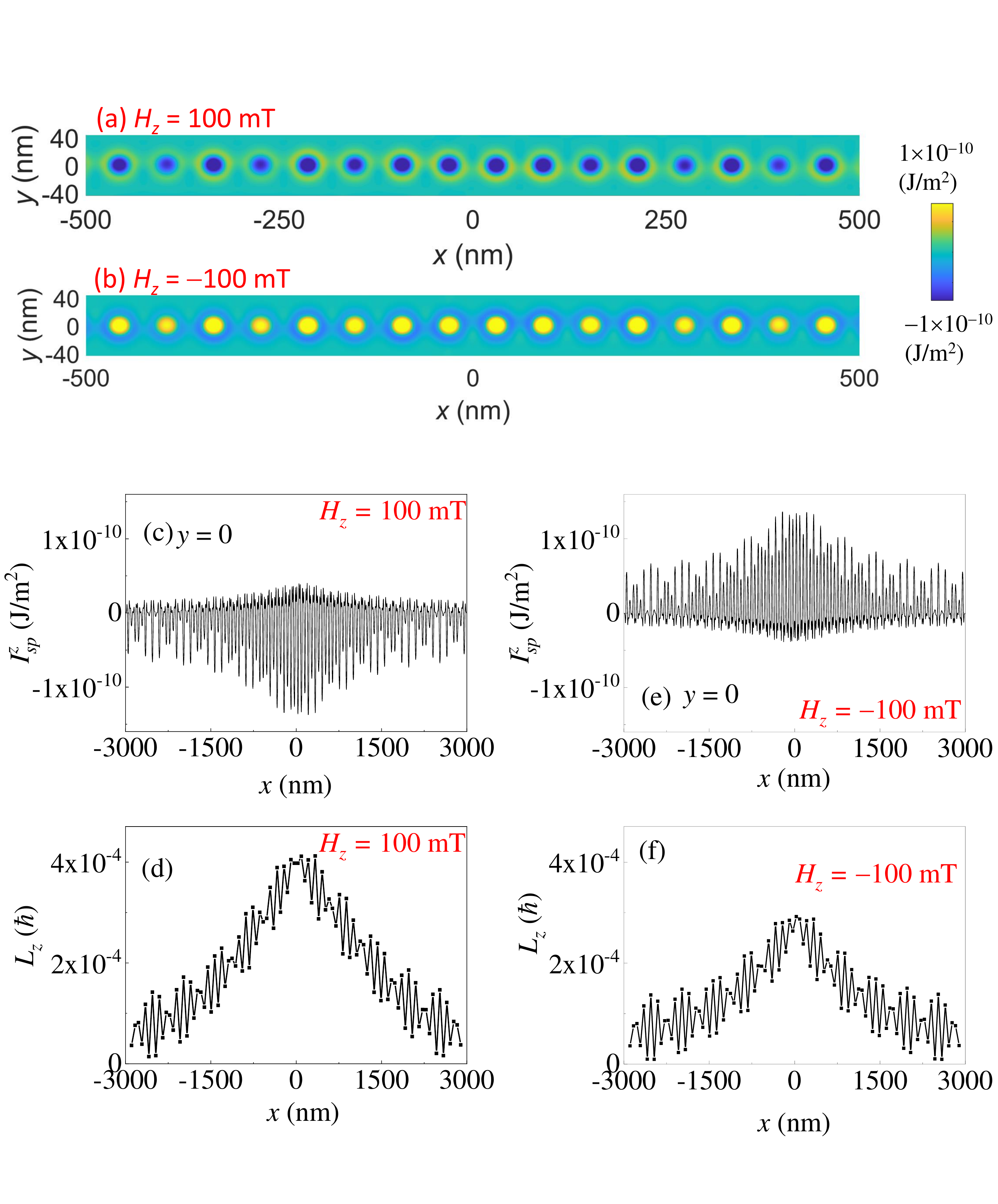}
  \caption{\label{pumping} Micromagnetic simulation for a single layer
    with 1D skyrmion lattice, stabilized by an applied external
    magnetic field $ H_z = 100 $ mT (a) and $ H_z = -100 $ mT
    (b). (c,e) The spatial profiles of the $ z $ component of spin
    pumping current $ I_{sp}^z(x, y=0)$, excited by a microwave field
    (applied near the region $ x = 0 $) with the frequency 0.3
    GHz. The spin pumping currents in the skyrmion center and at its
    boundary are opposite. This follows from magnetic structure of the
    skyrmion -- the magnetization in the skyrmion center is opposite
    to that at its boundary. (d, f) The $ z $ component of the orbital
    angular momentum $L_z$ of skyrmions due to phason excitations (the
    points correspond to the centers of skyrmions).}
\end{figure}

The collective phason excitations of the SkX can pump magnonic spin
current $ \vec{I}_{sp}$ into the adjacent metal. To calculate the
pumped spin current we exploit the formula
$ \vec{I}_{sp} = \frac{\hbar g_r}{4 \pi} \vec{m} \times \frac{\partial
  \vec{m}}{\partial t} $
(where $g_r$ is the real part of the dimensionless spin-mixing
conductance, assumed $g_r = 7 \times 10^{18}$ m$^{-2}$) and excite the
phason mode with the low frequency, equal to 0.3 GHz, microwave field
applied in the vicinity of the region $ x = 0 $. The spatial profile
of the $z$ component of the magnonic spin pumping current $ I_{sp}^z $
is shown in Fig.~\ref{pumping}(a) for positive magnetic field and in
the corresponding cross-section at $y=0$ in Fig.~\ref{pumping}(c).
The negative pumping current is mainly localized inside the skyrmion
lattice and propagates away from the excitation region ($ x = 0
$).
Outside the skyrmion region, the current $ I_{sp}^z $ becomes
positive.  In the absence of SkX, the magnetization oscillation with
frequency 0.3 GHz cannot propagate through the magnetic layer due to
the energy gap in the spectrum and therefore the pumping current
disappears. When reversing the direction of applied magnetic field,
$ H_z = -100 $ mT, and also of the magnetization direction, then the
spin current $ I_{sp}^z $ also changes its orientation as shown in
Fig.~\ref{pumping}(b,e).

The collective phason excitations in the SkX carry an orbital angular
momentum created by the dynamics of the three coupled helices. The
$z$-component of the orbital angular momentum can be calculated using
Noether's theorem \cite{PhysRevB.88.144413,PhysRevLett.124.217204} as
$L_z=(\hbar /S)\int l_z(x,y)\, dx\, dy$, where
$l_z(x,y)=m_z(\textbf{r}\times\vec\nabla\phi)_z$ is the orbital
angular momentum density, $S$ is the integration area and
$\phi=\arctan(m_y/m_x)$. The results of the calculations are shown in
Fig.~\ref{pumping}(d,f). Contrary to the magnonic spin pumping
current, the orbital angular momentum density does not change sign
upon the magnetic field reversal ($ H_z = \pm 100 $ mT).  We note that
the orbital angular momentum of the phason modes in the SkX can be
expressed in terms of the pseudo-Poynting vector
\cite{jia2019twisted}
as follows:
$\textbf{L}=(1/S)\int\textbf{r}\times \vec P\; dx dy$,
where
$ P_\mu=\frac{\hbar}{2}
\lbrace\tilde{E}^*(\partial_\mu\tilde{B})+\tilde{B}^*
(\partial_\mu\tilde{E})\rbrace $,
and we introduced the notations $\tilde{E}=m_x+im_y$ and
$\tilde{B}=i(m_x-im_y)$.
 The magnon attenuation effect leads to the spatial decay of $ L_z $. Therefore, when rescaling the orbital angular momentum dividing it by the magnon density $ n $, one can eliminate the effect of spatial decay and achieve  the quantized value of  $ L_z / n $,  $ L_z / n \approx \hbar$.

\subsection{The SkX in a SAF}

\begin{figure}
  \includegraphics[width=\columnwidth]{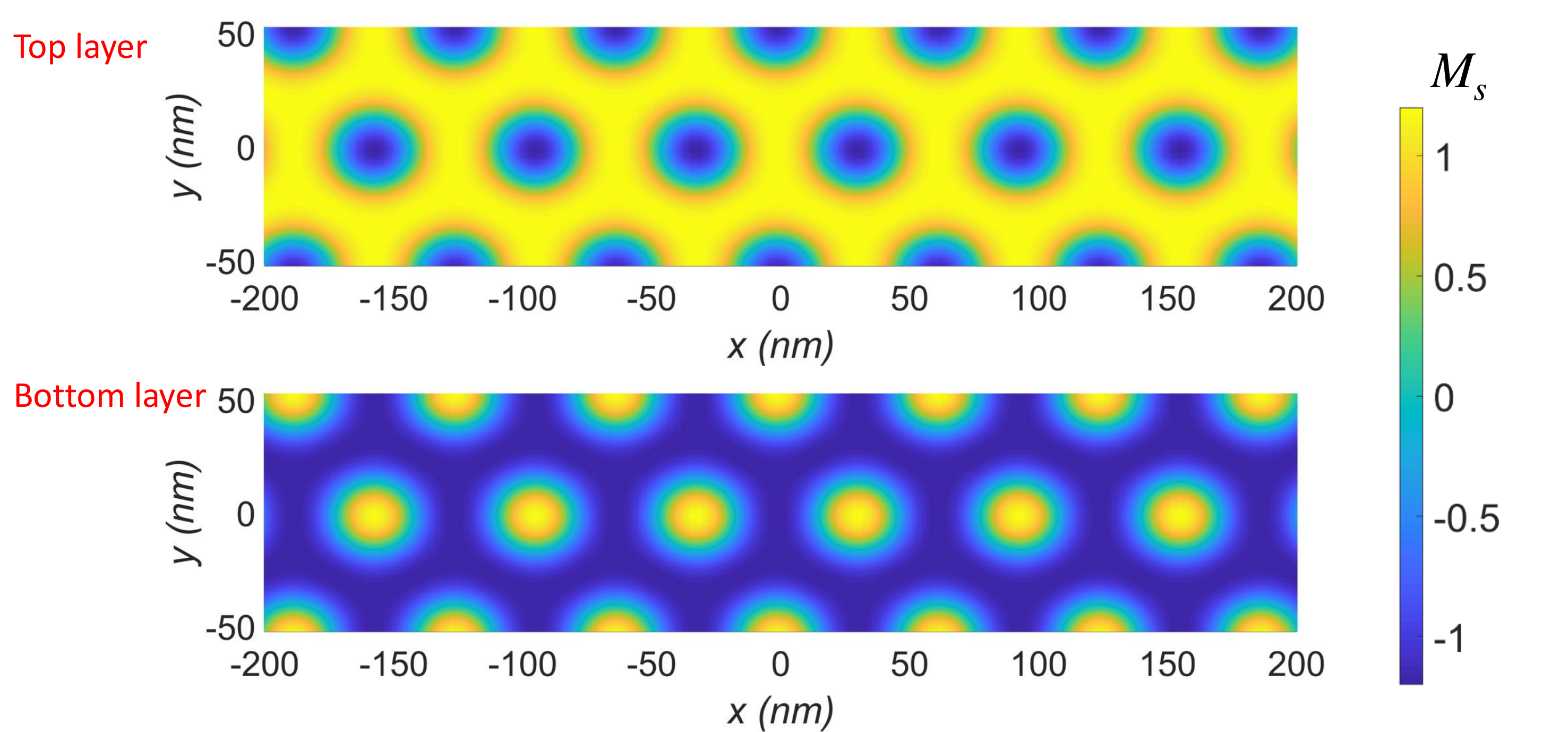}
  \caption{\label{modeldd} Schematics of the two-dimensional
    antiferromagnetic the SkX in a SAF. Both, top and bottom layers are
    shown.}
\end{figure}

\begin{figure}
  \includegraphics[width=\columnwidth]{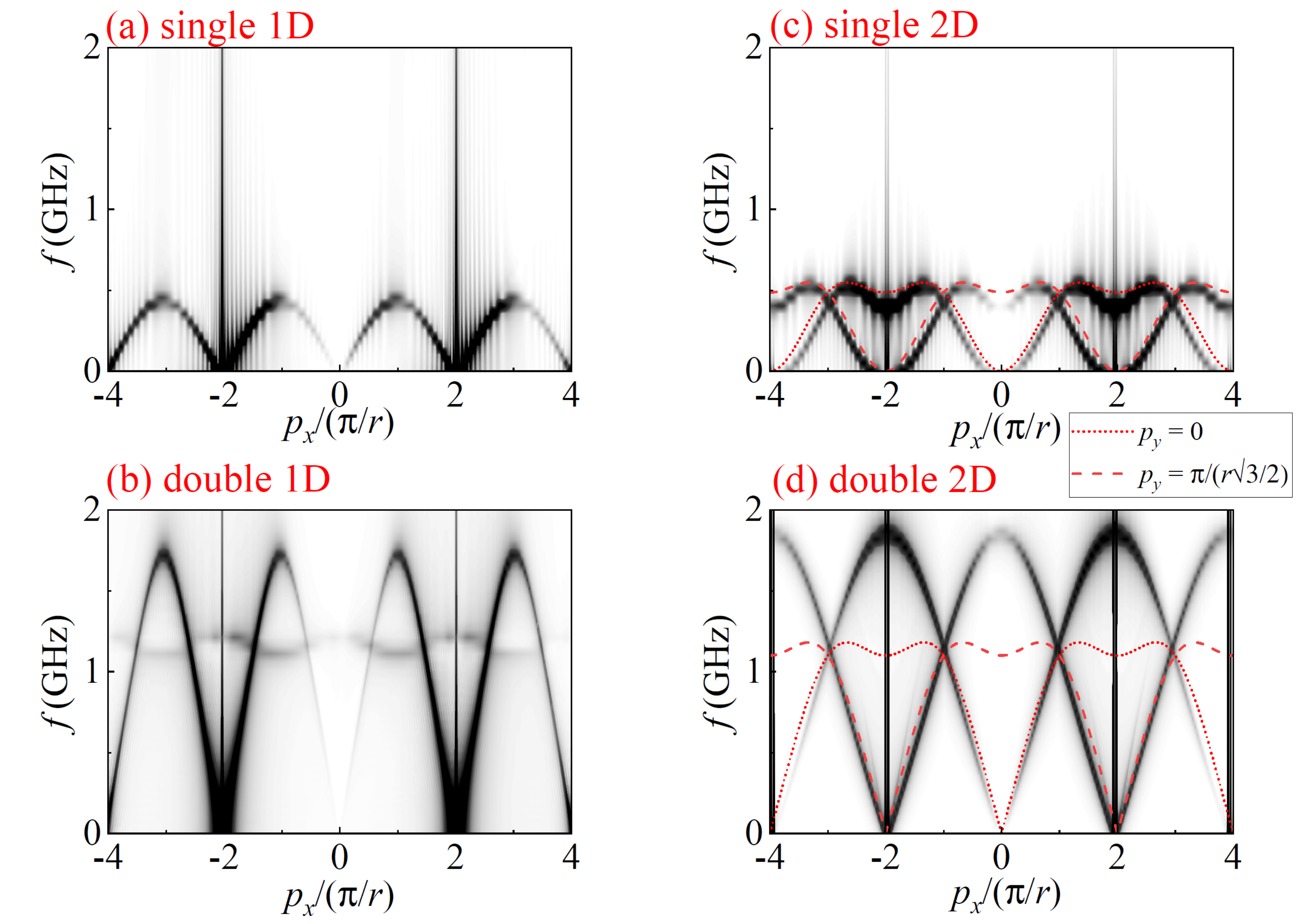}
  \caption{\label{dispersiondd} The spectrum of magnetic excitations in the
    low-frequency regime plotted for the single ferromagnetic layer
    (a,c) and for the SAF (b,d). The corresponding skyrmion lattices are
    1D (a,b) and 2D (c,d).  The red dotted ($ q_y = 0 $) and dashed  ($ q_y / (\pi/r) = 2/\sqrt{3} $) lines are calculated using Eqs. (\ref{fre}) and (\ref{doublefre}), respectively, for $ \sigma = 0.06 $ GHz and $ \sigma_{12} = 1 $ GHz.}
\end{figure}

Finally, we consider the skyrmion lattice (1D and 2D) in a SAF. The
corresponding magnetization profile in 2D case is plotted in
Fig.~\ref{modeldd}. The numerical results on the magnetic dynamics in
SAF are compared in Fig.~\ref{dispersiondd} with those for a single
magnetic layer. Both 1D and 2D cases are shown there.
The dispersion curves in the low frequency regime in the 1D SkX are
linear in both single layer and the SAF.  In turn, for 2D SkX, the
dispersion curve is still linear in the SAF but becomes quadratic in
the single layer. These results are consistent with those obtained
in section 2 for 2D systems within the model based on three coupled helices.
Moreover, they are also consistent with those obtained from Thiele's equations as will be discussed later.
The differences between the dispersion curves of 1D and 2D cases (especially for a single layer) originate from different boundary conditions. For 1D skyrmion lattice, we adopt a finite geo-boundary in the numerical calculations, and the boundary effect blocks one of the  degrees the freedom. This blocking is irrelevant in 2D case, where the periodic boundary conditions are employed.

\begin{figure}
  \includegraphics[width=\columnwidth]{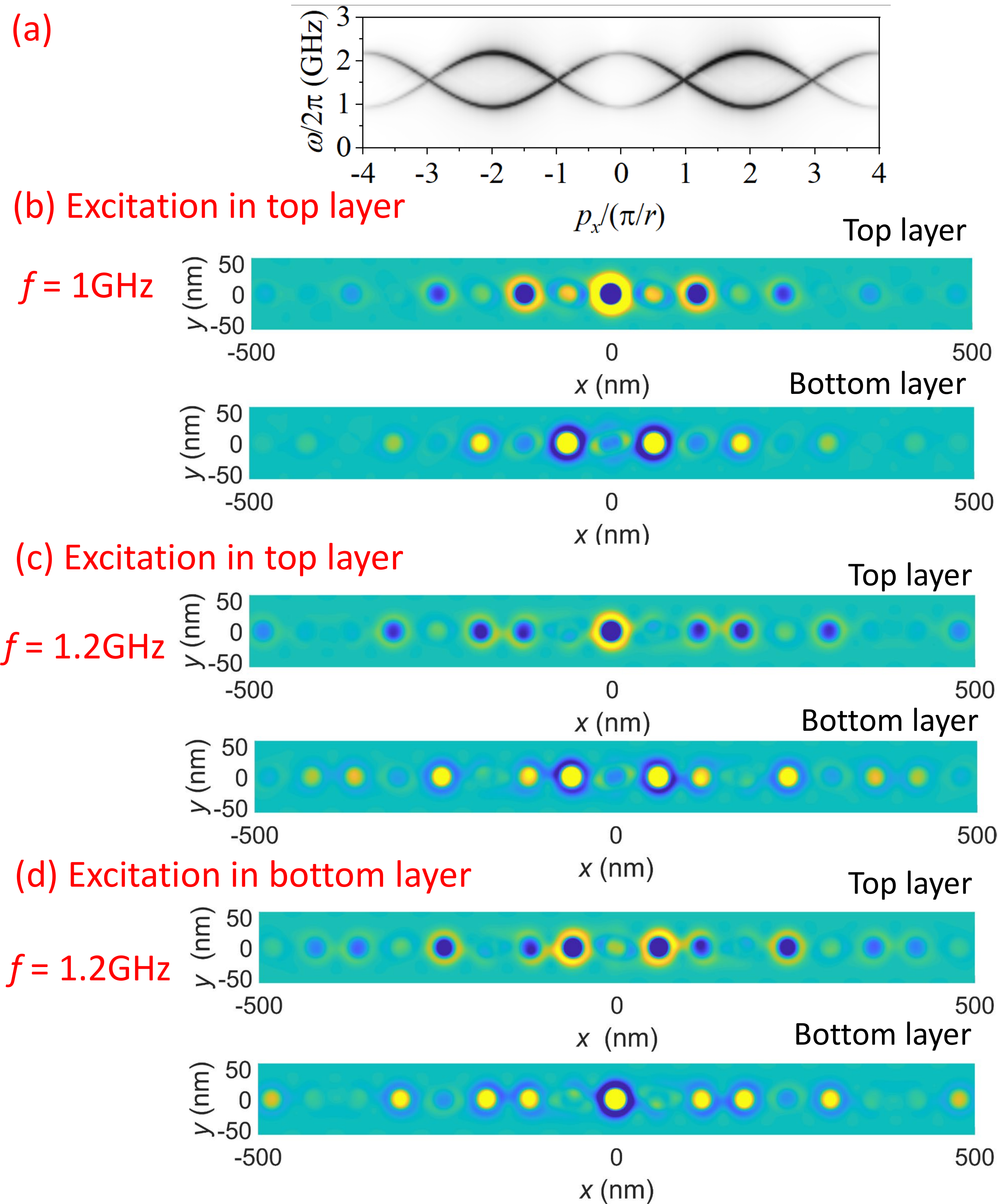}
  \caption{\label{pumping2} (a) The spectrum of 1D skyrmion lattice in a
    SAF when a bias field of 20 mT is applied near the skyrmion
    centers in both top and bottom layers.  (b-d) The spatial profiles
    of the $ z $ component of spin pumping current $ I_{sp}^z $ in the
    SAF layer. The spin pumping current and skyrmion precession are
    excited by the microwave field with frequencies 1 GHz (b) and 1.2
    GHz (c). The skyrmion precession is excited in the region
    $ x = 0 $ in the top layer. (d) The current $ I_{sp}^z $ generated
    by the skyrmion excitation in the bottom layer. The microwave
    field is applied in the region $ x = 0 $.}
\end{figure}

\begin{figure}
  \includegraphics[width=\columnwidth]{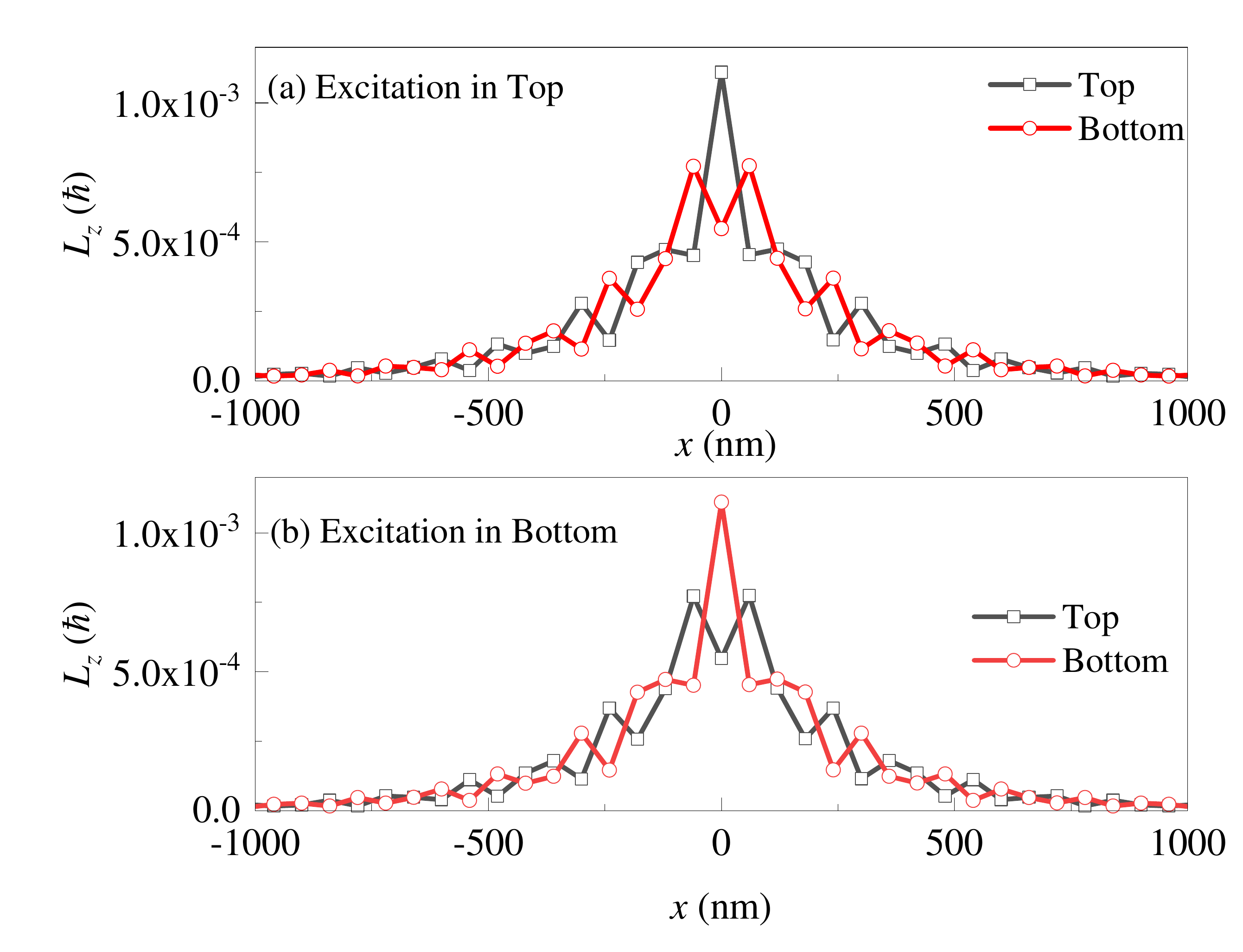}
  \caption{\label{orbitd} The $ z $ component of the orbital angular
    momentum $ L_z $ calculated for the 1D SkX in a SAF when the
    oscillation is excited by a microwave field applied near $ x = 0 $
    in (a) top layer and (b) bottom layer. The frequency of the field
    is 1.2 GHz.}
\end{figure}

The spin pumping current $ I_{sp}^z $ in the 1D SkX in a SAF is shown in
Fig.~\ref{pumping2}.  The bias field of 20 mT is applied near the
skyrmion centers in both top and bottom layers. This field plays the
role of a pinning potential that shifts upward the spectrum of the
system (Fig.~\ref{pumping2}(a)).  Applying the microwave field in the
vicinity of $ x = 0 $ in the top layer of SAF, one can induce the
skyrmion precession in both layers of the SAF. The skyrmion precession
in the top layer generates the negative spin pumping current
$ I_{sp}^z $ shown in blue color in Fig.~\ref{pumping2}(b).  Due to
the AFM coupling, magnetization dynamics in both layers are
correlated, and spin pumping current in the bottom layer is also
negative. However, the skyrmion precession and the magnitude of the
current are smaller as compared to those in the top layer.  The
negative $ I_{sp}^z $ in the SAF is similar to the skyrmion precession and
spin pumping current generated in a single layer
Fig.~\ref{pumping}(a). The microwave field applied to the bottom layer
Fig.~\ref{pumping2}(d) generates the skyrmion precession and positive
spin pumping current $ I_{sp}^z $ (shown in yellow color). The effect
is similar to the single-layer case Fig.~\ref{pumping}(b). Due to the
AFM coupling with a top layer, spin pumping current in the top layer
$ I_{sp}^z $ is also positive, but the current is smaller than that in
the bottom layer.

The interesting feature is the spatially non-uniform distribution of
the current. In particular, from Fig.~\ref{pumping2}(b) follows, that
at a certain distance from the $ x = 0 $ point, the spin pumping current
$ I_{sp}^z $ becomes positive, while at larger distances it again
switches the sign. Thus, we observe a spatially periodic switching of
the sign of current $ I_{sp}^z $.  The spatial distribution of the
current also depends on the frequency of the field,
Fig.~\ref{pumping2}(c).  The skyrmion precession in the bottom layer,
Fig.~\ref{pumping2}(d), switches the sign of current $ I_{sp}^z $ in
both layers as compared with Fig.~\ref{pumping2}(b).

The spatial alternation of the magnonic current can be explained as follows: Due to the antiferromagnetic coupling, both left-hand and right-hand precessions (concerning the local magnetization) coexist in the SAF. For the skyrmion with negative magnetization in the center and positive magnetization at the boundary, the pumping current $ I_{\rm sp}^z $ is negative in the center and positive at the boundary. When current  $ I_{\rm sp}^z $  reaches the neighboring skyrmion, it becomes negative again because the skyrmion permits only right-hand precession. In SAF, due to the coexisting left and right precessions, the current $I_{sp}^z$ can be either positive or negative. However, there is a significant asymmetry -- the right-hand precession is always stronger. Therefore, the $-z$ magnetization in the top layer and the right-hand precession induces the negative current, stronger than the negative current induced in the bottom layer due to the left-hand precession and positive $z$ magnetization. In Fig.~\ref{pumping2}(b), the negative current induced in the top layer leads to a smaller negative current in the bottom layer. The negative current emitted to the border of the first skyrmion changes the sign, and the positive current reaches the region of the second skyrmion. The positive $I_{sp}^z>0$ generates a stronger positive current in the bottom layer. Then $I_{sp}^z$ again becomes negative in the region of the third skyrmion, and the process is repeated further.

The orbital angular momentum density $ L_z $ of the 1D SkX in SAF is
plotted in Fig.~\ref{orbitd} for two cases: when the microwave field
is applied to the top (a) and bottom (b) layers of the SAF.  As one
can see, when the microwave is applied to the top layer, the orbital
angular momentum density in the top layer is larger and vice versa,
when the bottom layer is excited by the microwave field, the orbital
angular momentum density is larger in the bottom layer.

\section{Model based on Thiele's equations}

In case of skyrmions stabilized by the DMI, the center-of-mass motion of
an individual skyrmion can be described by Thiele's
equation~\cite{PhysRevLett.30.230}
\begin{equation}
\label{eq:1}
- \mathbf{G}\times \partial_t{\mathbf{r}} - \alpha \tensor{\cal{D}} \, \partial_t{\mathbf{r}} + {\mathbf F} = 0,
\end{equation}
where $\mathbf{G} = 4\pi N_{\mathrm{sk}}\, \hat{z}$ is the
gyrocoupling vector defined by the skyrmion topological charge
$N_{\mathrm{sk}} = \pm1$ and the unit vector $\hat{z}$ along the $z$ axis,
$ \tensor{\cal{D}}$ stands for a tensor of dissipative force, $\alpha$
is the Gilbert damping constant, and ${\bf F}$ is a force acting on
the skyrmion (${\bf F}=- \mathbf{\nabla} V $, with $V$ standing for
the corresponding potential energy).  The tensor $ \tensor{\cal{D}}$
 has the following form: $ \tensor{\cal{D}}_{ij} = D$ for
$(i,j) = (x,x)$ and $(i,j)=(y,y)$, while $\tensor{\cal{D}}_{ij} = 0$
otherwise \cite{Seidel}.  This particle-like description of skyrmion
dynamics is also valid for systems of interacting
skyrmions~\cite{Nagaosa2017,Martinez2016,Seidel}.

\subsection{Single layer case}

We consider first dynamical states od SkX in a single ferromagnetic
layer. To do this we model SkX as a periodic (in equilibrium)
array of coupled skyrmions confined in the
position ${\bf R}_i$, where each skyrmion  is surrounded by  six nearest neighbours. In a nonequilibrium (dynamical) state, position
of the $i$th skyrmion, ${\bf r}_i$ can written as
${\bf r}_i = {\bf R}_i+{\bf u}_i$, where ${\bf u}_i$ stands for a
deviation of the skyrmion center from its equilibrium position.

Let us consider two nearest-neighbour interacting skyrmions, $i$ and $j$,  located in equilibrium at the positions ${\bf R}_i$ and ${\bf R}_j$, and let $\bf{e}_{ij}$ stands for a unit vector along  ${\bf R}_j-{\bf R}_i$. Energy of such skyrmions in nonequilibrium is generally anisotropic, i.e. it depends on relative orientation of the skyrmion displacements and the vector  $\bf{e}_{ij}$. Thus, the relative displacement, $ \mathbf{u}_{i} -
\mathbf{u}_{j}$, can be decomposed into the component $[\bf{e}_{ij}\cdot (\mathbf{u}_{i} -\mathbf{u}_{j})] \bf{e}_{ij}$ along the vector $\bf{e}_{ij}$ and the component $(\mathbf{u}_{i} -\mathbf{u}_{j}) -[\bf{e}_{ij}\cdot (\mathbf{u}_{i} -\mathbf{u}_{j})] \bf{e}_{ij}$ normal to  $\bf{e}_{ij}$.

Since the total force acting on the $i$th skyrmion
is a superposition of the forces from its nearest neighbours,  we write the Thiele's equation  in the following form:
\begin{eqnarray}
- \mathbf{G} \times \partial_{t} \mathbf{u}_{i} - \alpha  \tensor{\cal{D}} \, \partial_{t}  \mathbf{u}_{i} = \sigma_{\parallel} \sum_{ \langle j \rangle }  [\bf{e}_{ij}\cdot (\mathbf{u}_{i} -\mathbf{u}_{j})] \bf{e}_{ij}  \nonumber \\+  \sigma_{\perp} \sum_{ \langle j \rangle }\{ (\mathbf{u}_{i} -\mathbf{u}_{j}) -[\bf{e}_{ij}\cdot (\mathbf{u}_{i} -\mathbf{u}_{j})] \bf{e}_{ij}\},
\end{eqnarray}
where $ \sigma_{\parallel} $ and $ \sigma_{\perp} $ are the two coupling parameters
and $ \langle j \rangle $ denotes the summation over
the nearest-neighbour skyrmions. A particular skyrmion (say the $i$th one) is surrounded by  six nearest neighbours indexed with $j=1$ to $j=6$.
They  correspond to ${\bf e}_{i1}=(1,0)$, ${\bf e}_{i2}=(1/2,\sqrt{3}/2)$, ${\bf e}_{i3}=(-1/2,\sqrt{3}/2)$, ${\bf e}_{i4}=(-1,0)$,  ${\bf e}_{i5}=(-1/2,-\sqrt{3}/2)$, and ${\bf e}_{i6}=(1/2,-\sqrt{3}/2)$.

In the following we assume skyrmions
corresponding to $N_{sk}=1$ and adequately normalize  $ \tensor{\cal{D}}$ as well as $\sigma_\parallel$ and $\sigma_\perp$.
Taking into account the explicit form of the tensor
$ \tensor{\cal{D}}$~\cite{Martinez2016,Seidel}, the Thiele's equations
for a skyrmion lattice in a single ferromagnetic layer can be written in
the form
%
%
%
\begin{eqnarray} 
- \partial_{t} u_i^y - \alpha {\cal{D}}\, \partial_{t} u_i^x
=\sum_{\langle j \rangle } \{ \left[ \sigma_{\parallel} e^2_{ijx}+\sigma_{\perp} (1-e^2_{ijx}) \right]
\nonumber \\
\times (u_i^x - u_j^x)
  + (\sigma_{\parallel}-\sigma_{\perp})e_{ijx}e_{ijy}(u_i^y  - u_j^y)\},
\end{eqnarray}

\begin{eqnarray} 
 \partial_{t} u_i^x - \alpha {\cal{D}}\, \partial_{t} u_i^y
=\sum_{\langle j \rangle } \{ \left[ \sigma_{\parallel} e^2_{ijy}+\sigma_{\perp} (1-e^2_{ijy}) \right]
\nonumber \\
\times (u_i^y - u_j^y)
  + (\sigma_{\parallel}-\sigma_{\perp})e_{ijx}e_{ijy}(u_i^x - u_j^x)\}.
\end{eqnarray}

We look now for solutions of the above equations in the Bloch's form,
$ u_i^{x(y)} = u_{0}^{x(y)} e^{i [{\bf{p}} \cdot {\bf{R}}_i -
  \omega (t)]}$. For simplicity, we neglect the term proportional to ${\cal{D}}$ as being small due to a small damping parameter $\alpha$. Then, taking into account positions of all six neighbours, one finds  from  Eqs (5) and (6) the following equations:
\begin{widetext}%
\begin{equation}
i\omega  u_{0}^y = \{2\sigma_\parallel [1-\cos (p_x\delta )]   +(\sigma_\parallel +3\sigma_\perp ) [1-f_1({\bf {p}})] \}u_0^x +
\sqrt{3}(\sigma_\parallel -\sigma_\perp ) f_2 ({\bf{p}}) u_0^y ,
\end{equation}
\begin{equation}
-i\omega  u_{0}^x = \{2\sigma_\perp [1-\cos (p_x\delta )]   +(3\sigma_\parallel +\sigma_\perp ) [1-f_1({\bf {p}})] \}u_0^y +
\sqrt{3}(\sigma_\parallel -\sigma_\perp ) f_2 ({\bf{p}}) u_0^x ,
\end{equation}
\end{widetext}
where
\begin{eqnarray}
f_1({\bf {p}})=\cos (\frac{p_x\delta}{2}) \cos (\frac{\sqrt{3}}{2}p_y\delta ),  \\
f_2({\bf {p}})=\sin (\frac{p_x\delta}{2}) \sin (\frac{\sqrt{3}}{2}p_y\delta ),
\end{eqnarray}
and $\delta$ is the distance between nearest-neighbour skyrmions.
Defining  the column vector ${\bf{z}}= (u_0^{x}, u_0^{y})^T$, the above equations (16) and (17) can be written as
\begin{equation}
\hat{H}{\bf{z}} =0,
\end{equation}
where the $2\times 2$ matrix $\hat{H}$ has the form
\begin{equation}
\hat{H}=
\begin{pmatrix}
B & A-i\omega \\
A+i\omega & B^\prime\\
\end{pmatrix}
\end{equation}
with
\begin{eqnarray}
&A=\sqrt{3}(\sigma_\parallel -\sigma_\perp ) f_2 ({\bf{p}}) ,\\
&B=2\sigma_\parallel [1-\cos (p_x\delta )]   +(\sigma_\parallel +3\sigma_\perp ) [1-f_1({\bf {p}})] , \\
&B^\prime = 2\sigma_\perp [1-\cos (p_x\delta )]   +(3\sigma_\parallel +\sigma_\perp ) [1-f_1({\bf {p}})] .
\end{eqnarray}

The condition of vanishing determinant of the matrix $\hat{H}$ leads to the following frequency $\omega$:
\begin{eqnarray}
\omega =  \left[\{2\sigma_\parallel [1-\cos (p_x\delta )]   +(\sigma_\parallel +3\sigma_\perp ) [1-f_1({\bf {p}})] \}\right. \nonumber \\
\left.\times \{2\sigma_\perp [1-\cos (p_x\delta )]   +(3\sigma_\parallel +\sigma_\perp ) [1-f_1({\bf {p}})] \}\nonumber \right. \\
\left. - 3 [(\sigma_\parallel -\sigma_\perp )  f_2 ({\bf{p}})]^2\right]^{1/2} .\hspace{0.5cm}
\end{eqnarray}
In the limit of small $ |\vec{p}| $, the frequency can expanded as $ \omega \approx  \frac{3\sqrt{(3\sigma_{\|}+\sigma_{\perp})(\sigma_{\|}+3\sigma_{\perp})}}{8} \mathbf{p}^2 r^2 $.

In the isotropic limit, $\sigma_\parallel =\sigma_\perp =\sigma$, this formula reduces to the following simple expression for the phason frequency $\omega$:
\begin{equation}
\label{fre}
\omega = 2\sigma [3-\xi ({\bf {p}}) ],
\end{equation}
where $\xi ({\mathbf{p}})$ is defined as
\begin{equation}
\xi ({\mathbf{p}})= \cos(p_{x} \delta) + 2 \cos\left(\frac{\sqrt{3}}{2} p_{y}
  \delta\right)\cos(\frac{1}{2} p_{x}\delta ).
\end{equation}
From this formula one can easily note  that the spectrum is gapless. Moreover, from expansion with respect to $p$ follows that the spectrum in the small wavevector limit is quadratic in $p$, i.e. $\omega \sim  p^2$, as is proved by the series expansion $ \omega \approx  \frac{3 \sigma}{2} \mathbf{p}^2 r^2 $.

\subsection{Thiele's equations for SAF}

Now, we apply the description based on the Thiele's equations to the
SkX in SAF.  Let the skyrmions in the top (FM1) layer (see
Fig. \ref{model}) are in the positions
${\bf r}_{i}^{(1)} = {\bf R}_{i}+{\bf u}_{i}^{(1)}$, while in the bottom
layer are in positions ${\bf r}_{i}^{(2)} = {\bf R}_{i}+{\bf u}_{i}^{(2)}$. We assumed here that in equilibrium skyrmions in both layers are in the same positions,  ${\bf R}_{i}^{(1)}= {\bf R}_{i}^{(2)} = {\bf R}_{i}$.
Accordingly, the Thiele's equations can be written as
\begin{widetext}
\begin{eqnarray}
- \mathbf{G}^{(1)} \times \partial_{t} \mathbf{u}_{i}^{(1)} - \alpha^{(1)}  \tensor{\cal{D}}^{(1)} \, \partial_{t}  \mathbf{u}_{i}^{(1)} = \sigma_{1\parallel} \sum_{ \langle j \rangle }  [\bf{e}_{ij}\cdot (\mathbf{u}_{i}^{(1)} -\mathbf{u}_{j}^{(1)})] \bf{e}_{ij}   \nonumber \\
+  \sigma_{1\perp} \sum_{ \langle j \rangle }\{ (\mathbf{u}_{i}^{(1)} -\mathbf{u}_{j}^{(1)}) -[\bf{e}_{ij}\cdot (\mathbf{u}_{i}^{(1)} -\mathbf{u}_{j}^{(1)})] \bf{e}_{ij}\} + \sigma_{\rm 12} (\mathbf{u}_{i}^{(1)} - \mathbf{u}_{i}^{(2)}),
\end{eqnarray}
\begin{eqnarray}
- \mathbf{G}^{(2)} \times \partial_{t} \mathbf{u}_{i}^{(2)} - \alpha^{(2)}  \tensor{\cal{D}}^{(2)} \, \partial_{t}  \mathbf{u}_{i}^{(2)} = \sigma_{2\parallel} \sum_{ \langle j \rangle }  [\bf{e}_{ij}\cdot (\mathbf{u}_{i}^{(2)} -\mathbf{u}_{j}^{(2)})] \bf{e}_{ij}   \nonumber \\
+  \sigma_{2\perp} \sum_{ \langle j \rangle }\{ (\mathbf{u}_{i}^{(2)} -\mathbf{u}_{j}^{(2)}) -[\bf{e}_{ij}\cdot (\mathbf{u}_{i}^{(2)} -\mathbf{u}_{j}^{(2)})] \bf{e}_{ij}\} + \sigma_{\rm 12} (\mathbf{u}_{i}^{(2)} - \mathbf{u}_{i}^{(1)}),
\end{eqnarray}
\end{widetext}
where the in-plane coupling constants in the top (bottom)  layer  are $\sigma_{1\parallel}$ and $\sigma_{1\perp}$ ($\sigma_{2\parallel}$ and $\sigma_{2\perp}$), whereas  $\sigma_{\rm 12}$ denotes the  coupling parameter between skyrmions in different layers (each skyrmion in one layer has only a single nearest-neighbour skyrmion in the second layer). Note, the coupling between skyrmions in two different layers is isotropic.  Other parameters in the two layers are distinguished with the upper indices  $(1)$ and $(2)$.

Due to antiferromagnetic interlayer coupling between
layers, the magnetic texture in the ferromagnetic layer
FM2 is topologically reversed to that in the layer FM1,
and consequently topological charges of skyrmions in
these layers are also opposite, $N_{\rm sk}^{(2)} = -N_{\rm sk}^{(1)}$. Accordingly, the Thiele's equations
can be written explicitly as follows:
\begin{widetext}
\begin{eqnarray} 
- \partial_{t} u_{iy}^{(1)} - \alpha^{(1)} {\cal{D}}^{(1)}\, \partial_{t} u_{ix}^{(1)}
=\sum_{\langle j \rangle } \{ \left[ \sigma_{1\parallel} e^2_{ijx}+\sigma_{1\perp} (1-e^2_{ijx}) \right]
(u_{ix}^{(1)} - u_{jx}^{(1)})\nonumber \\
  + (\sigma_{1\parallel}-\sigma_{1\perp})e_{ijx}e_{ijy}(u_{iy}^{(1)} - u_{jy}^{(1)})\} + \sigma_{12} (u_{ix}^{(1)} - u_{ix}^{(2)}) ,
\end{eqnarray}
\begin{eqnarray} 
 \partial_{t} u_{ix}^{(1)} - \alpha^{(1)} {\cal{D}}^{(1)}\, \partial_{t} u_{iy}^{(1)}
=\sum_{\langle j \rangle } \{ \left[ \sigma_{1\parallel} e^2_{ijy}+\sigma_{\perp} (1-e^2_{ijy}) \right]
(u_{iy}^{(1)} - u_{jy}^{(1)})\nonumber \\
  + (\sigma_{1\parallel}-\sigma_{1\perp})e_{ijx}e_{ijy}(u_{ix}^{(1)} - u_{jx}^{(1)})\} + \sigma_{12} (u_{iy}^{(1)} - u_{iy}^{(2)}) ,
\end{eqnarray}
\begin{eqnarray} 
+ \partial_{t} u_{iy}^{(2)} - \alpha^{(2)} {\cal{D}}^{(2)}\, \partial_{t} u_{ix}^{(2)}
=\sum_{\langle j \rangle } \{ \left[ \sigma_{2\parallel} e^2_{ijx}+\sigma_{2\perp} (1-e^2_{ijx}) \right]
(u_{ix}^{(2)} - u_{jx}^{(2)})\nonumber \\
  + (\sigma_{2\parallel}-\sigma_{2\perp})e_{ijx}e_{ijy}(u_{iy}^{(2)} - u_{jy}^{(2)})\} + \sigma_{12} (u_{ix}^{(2)} - u_{ix}^{(1)}),
\end{eqnarray}
\begin{eqnarray} 
- \partial_{t} u_{ix}^{(2)} - \alpha^{(2)} {\cal{D}}^{(2)}\, \partial_{t} u_{iy}^{(2)}
=\sum_{\langle j \rangle } \{ \left[ \sigma_{2\parallel} e^2_{ijy}+\sigma_{2\perp} (1-e^2_{ijy}) \right]
(u_{iy}^{(2)} - u_{jy}^{(2)})\nonumber \\
  + (\sigma_{2\parallel}-\sigma_{2\perp})e_{ijx}e_{ijy}(u_{ix}^{(2)} - u_{jx}^{(2)})\} + \sigma_{12} (u_{iy}^{(2)} - u_{iy}^{(1)}).
\end{eqnarray}
\end{widetext}

Similarly as in the single layer case,
we neglect the term proportional to ${\cal{D}}$ and look   for solutions of the above equations in the Bloch's wave form,
$ u_i^{x(y)} = u_{0}^{x(y)} e^{i [{\bf{p}} \cdot {\bf{R}}_i -
  \omega (t)]}$.  Then, taking into account positions of all six in-plane nearest neighbours,  from  Eqs (30-33) one finds,
\begin{widetext}%
\begin{equation}
i\omega  u_{0}^{(1)y} = \{2\sigma_{1\parallel} [1-\cos (p_x\delta )]   +(\sigma_{1\parallel} +3\sigma_{1\perp} ) [1-f_1({\bf {p}})] \}u_0^{(1)x} +
\sqrt{3}(\sigma_{1\parallel} -\sigma_{1\perp} ) f_2 ({\bf{p}}) u_0^{(1)y} +\sigma_{12}[u_{0}^{(1)x}-u_{0}^{(2)x}],
\end{equation}
\begin{equation}
-i\omega  u_{0}^{(1)x} = \{2\sigma_{1\perp} [1-\cos (p_x\delta )]   +(3\sigma_{1\parallel} +\sigma_{1\perp} ) [1-f_1({\bf {p}})] \}u_0^{(1)y} +
\sqrt{3}(\sigma_{1\parallel} -\sigma_{1\perp} ) f_2 ({\bf{p}}) u_0^{(1)x} +\sigma_{12}[u_{0}^{(1)y}-u_{0}^{(2)y}],
\end{equation}
and
\begin{equation}
-i\omega  u_{0}^{(2)y} = \{2\sigma_{2\parallel} [1-\cos (p_x\delta )]   +(\sigma_{2\parallel} +3\sigma_{2\perp} ) [1-f_1({\bf {p}})] \}u_0^{(2)x} +
\sqrt{3}(\sigma_{2\parallel} -\sigma_{2\perp} ) f_2 ({\bf{p}}) u_0^{(2)y} +\sigma_{12}[u_{0}^{(2)x}-u_{0}^{(1)x}],
\end{equation}
\begin{equation}
i\omega  u_{0}^{(2)x} = \{2\sigma_{2\perp} [1-\cos (p_x\delta )]   +(3\sigma_{2\parallel} +\sigma_{2\perp} ) [1-f_1({\bf {p}})] \}u_0^{(2)y} +
\sqrt{3}(\sigma_{2\parallel} -\sigma_{2\perp} ) f_2 ({\bf{p}}) u_0^{(2)x} +\sigma_{12}[u_{0}^{(2)y}-u_{0}^{(1)y}].
\end{equation}
\end{widetext}

Similarly as in the case of  single layer, we define  the column vector ${\bf{z}}= (u_0^{(1)x}, u_0^{(1)y}, (u_0^{(2)x}, u_0^{(2)y})^T$. Then, the above equations (34) to (37) can be written as
\begin{equation}
\hat{\tilde{H}}{\bf{z}} =0.
\end{equation}
Here, the $4\times 4$ matrix $\hat{\tilde{H}}$ takes the form
\begin{equation}
\hat{\tilde{H}}=
\begin{pmatrix}
\hat{H}^{(1)}+\sigma_{12}\hat{I} & -\sigma_{12}\hat{I}  \\
-\sigma_{12}\hat{I} & \hat{H}^{(2)*} + \sigma_{12}\hat{I}\\
\end{pmatrix}
\end{equation}
where $\hat{H}^{(1)}$ and $\hat{H}^{(2)}$ are the matrices $\hat{H}$ (see Eq.(21)) corresponding to the top and bottom layers, respectively, while $\hat{I}$ is the $/2\times 2$ unit matrix. Frequency can be determined numerically from the condition of vanishing determinant of  the matrix $\hat{\tilde{H}}$ (Eq.(39)).

\begin{figure}
  \includegraphics[width=\columnwidth]{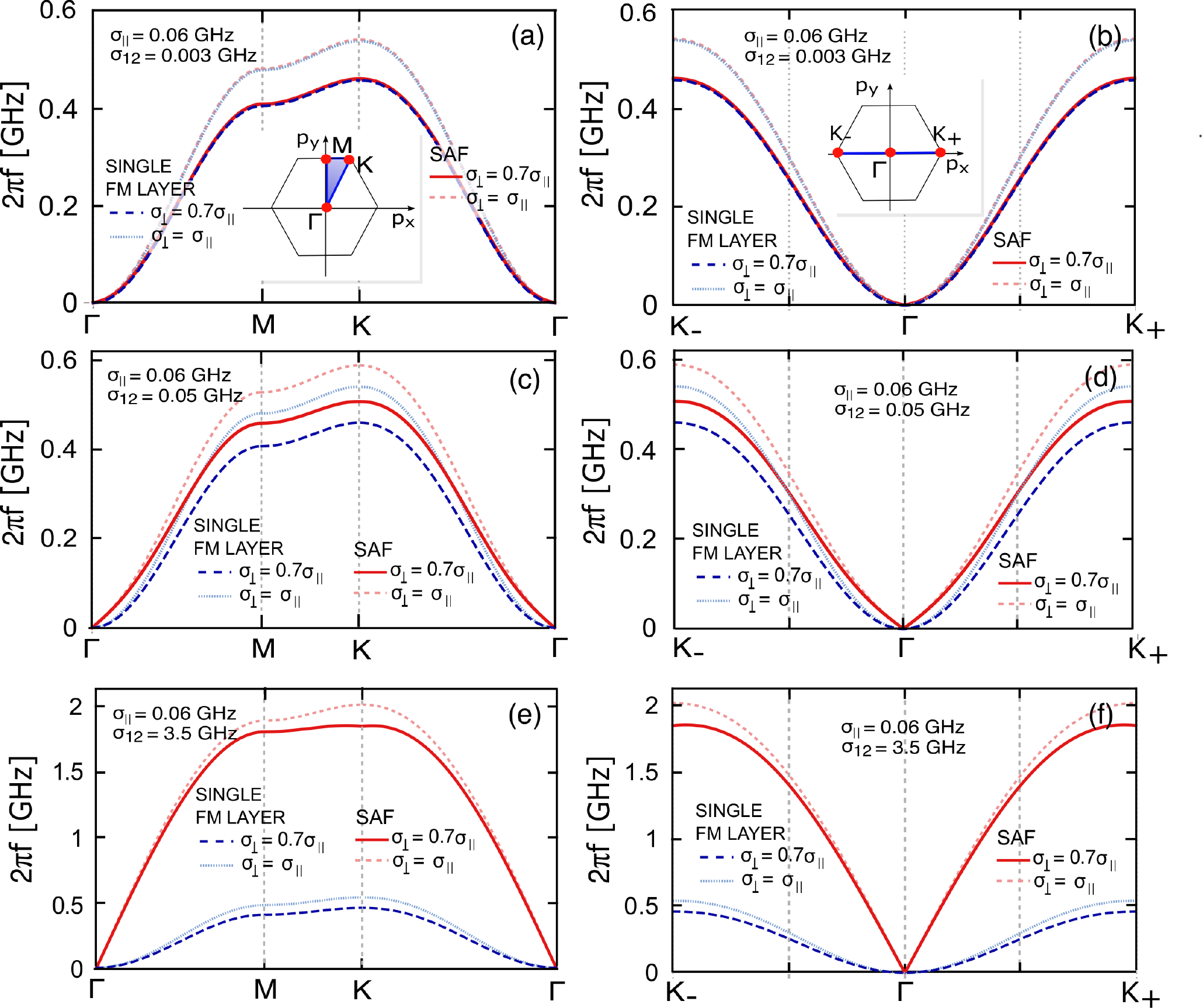}
  \caption{\label{fig:dispThiele}Dispersion relations of the phason
    excitations in 2D within the approach based on Thiele's equation, plotted  along
   the main crystal directions in the Brillouin zone of SkX. The
    dotted line corresponds to the modes in a single ferromagnetic
    layer, while the solid line presents the positive mode,
    $\omega_{+}(\vec{p})$ in a SAF. The negative mode in the SAF is exactly
    opposite to the positive one,
    $\omega_{-}(\vec{p})=-\omega_{+}(\vec{p})$, so it is not presented
    in the figure. The red curves correspond to a symmetric SAF ($\sigma_{1\parallel}=\sigma_{2\parallel}=\sigma_{\parallel}$ and $\sigma_{1\perp}=\sigma_{2\perp}=\sigma_{\perp}$) while the blue ones to a single layer. For each case two situations are distinguished:  isotropic ($\sigma_{\parallel}= \sigma_{\perp}$) and anisotropic $\sigma_{\parallel}> \sigma_{\perp}$).
    Different panels correspond to the following
    situations: (a,b) $\sigma_{12}\ll \sigma_\parallel$; (c,d)
    $\sigma_{12}\approx \sigma_\parallel$; and (e,f) $\sigma_{12}\gg \sigma_\parallel$. }
\end{figure}

In the isotropic  and symmetric case, $\sigma_{1\parallel}=\sigma_{1\perp}=\sigma_{2\parallel}=\sigma_{2\perp} = \sigma$,
one can find the following  simple analytical formula for the phason frequency
\begin{equation} \label{doublefre}
\omega_{\pm} = \pm 2 \sqrt{\left(3-\xi ({\mathbf{p}}) \right) \sigma \left[(3-\xi ({\mathbf{p}})) \sigma  + \sigma_{12}\right]},
\end{equation}
where $\xi ({\mathbf{p}})$ is defined by Eq.(27).
We note that the positive (negative) frequency corresponds to the
right-hand (left-hand) precession of the skyrmions in SAF, and
negative mode is opposite to the positive one,
$\omega_{-}({\bf p})=-\omega_{+}({\bf p})$.
Importantly, as follows from series expansion, the mode in SAF
becomes then linear with $\bf p$ in the limit of small
$|\vec{p}|$, $ \omega_{\pm} \sim \pm p$. This is proved by the small $ |\vec{p}| $ expansion, $ \omega_{\pm} \approx \pm \sqrt{2\sigma \sigma_{12}} |\vec{p}| r$. This linear in $|\vec{p}|$ dependence of the excitation frequency in
SAF is much more evident when $\sigma_{12} \gg \sigma$
(Fig.~\ref{fig:dispThiele} (e),(f)). Such a situation seems to be
physically more likely, as the in-plane skyrmion lattice constant is much larger than
the thickness of the nonmagnetic spacer layer in SAF.

In Fig.~\ref{fig:dispThiele} we present numerical results on the dispersion curves of the phason modes
propagating in a single ferromagnetic layer as well as in a symmetric  ($\sigma_{1\parallel}=\sigma_{2\parallel}=\sigma_{\parallel}$ and $\sigma_{1\perp}=\sigma_{2\perp}=\sigma_{\perp}$) SAF.  We also distinguish between  isotropic ($\sigma_{\parallel}= \sigma_{\perp}$) and anisotropic ($\sigma_{\parallel} > \sigma_{\perp}$) situations. When the coupling between skyrmions across the
spacer layer is weak, $\sigma_{12} \ll \sigma$, the dispersion
relation for a SAF is very similar to that found in the single
ferromagnetic layer (Fig.~\ref{fig:dispThiele} (a),(b)).  However,
when the interlayer coupling between the skyrmions is comparable to or stronger than
the intralayer one, $\sigma_{12} \simeq \sigma$, the mode in the SAF is
significantly different from that for a single layer
(Fig.~\ref{fig:dispThiele} (c),(d)).
The frequency of phason excitations also depends on the coupling anisotropy and becomes reduced when  $\sigma_\perp$ becomes smaller than $\sigma_\parallel$

 The results derived from the Thiele's equations are in agreement with those obtained from numerical simulations for 2D SkX's.  This is shown in
Fig.5(c,d), where  we have added the results obtained from the Thiele's, see the red dotted and dashed lines. The  agrement between analytical results derived from the Thiele's equations and those obtained from numerical simulations is satisfactory, and confirm that the low energy spectrum in a single SkX layer is quadratic while the spectrum of 2D SkX in SAF is  linear.

\section{Summary and conclusions}

In summary, we studied phason excitations of a skyrmion lattice in
  synthetic antiferromagnets, i.e., in two ferromagnetic layers coupled
  antiferromagnetically due to interlayer exchange interaction.
We have considered the magnetic dynamics of a skyrmion lattice in a SAF.
Three different methods have been used to analyze the
spectrum of magnetic dynamics: (i) an analytical approach based on the
presentation of the SkX as a superposition of three helices, (ii)
numerical micromagnetic simulations, and (iii) a simplified approach
based on the Thiele's equations. Interestingly, all the three
approaches give consistent results, i.e. all lead to gapless and
linear phason modes.

For comparison, we have also analysed  1D and 2D  single-layer skyrmion
lattices. In the 2D case we found gapless modes with quadratic dispersion in the small wavevevector limit. In turn, in the 1D case the situation is different, and we found linear behavior (except the approach based on the Thiele's equation, where this hehavior is quadratic).

We have also analyzed numerically the spin current pumped by skyrmion
dynamics (phasons). Spatial variation of the pumped current
polarization is shown to reveal internal magnetic texture of the
skyrmions. In addition, orbital angular momentum associated with
phason dynamics induced externally by a microwave field has also been
calculated.

We found that three independent helix modes hosted in the synthetic
antiferromagnetic material possess beneficial features for generation of magnonic spin currents and
implementation in spin caloritronics.

\textbf{Acknowledgments} This work was supported by the National
Science Center in Poland as a research Project No. DEC-2017/27/B/ST3/
02881 (VKD), and by the Norwegian Financial Mechanism 2014-2021 under
the Polish-Norwegian Research Project NCN GRIEG (2Dtronics)
no. 2019/34/H/ST3/00515 (AD,JB). It was also supported by the DFG
through the SFB-TRR 227, Shota Rustaveli National Science Foundation
of Georgia (SRNSFG) (Grant No. FR-19-4049), the National Natural
Science Foundation of China (Grants No. 12174452, No. 12074437, No. 11704415), the Natural Science
Foundation of Hunan Province of China (Grants No. 2022JJ20050, No. 2021JJ30784), and Grant-in-Aid
for Scientific Research (B) (No. 17H02929) from the Japan Society for
the Promotion of Science.  A. E. acknowledges financial support from
DFG through priority program SPP1666 (Topological Insulators),
SFB-TRR227, and OeAD Grants No. HR 07/ 2018 and No. PL 03/2018.

\bibliography{./2nems}

\end{document}